\documentclass{elsarticle}

\usepackage{amssymb}
\usepackage{braket}
\usepackage{hyperref}

\journal{Physica A}

\begin{document}

\begin{frontmatter}

\title{On the relationship between the Hurst exponent, the ratio of the mean square successive difference to the variance, and the number of turning points}

\author{Mariusz Tarnopolski}
\address{Astronomical Observatory, Jagiellonian University, Orla 171, PL-31-244 Krak\'ow, Poland}
\ead{mariusz.tarnopolski@uj.edu.pl}

\begin{abstract}
The long range dependence of the fractional Brownian motion (fBm), fractional Gaussian noise (fGn), and differentiated fGn (DfGn) is described by the Hurst exponent $H$. Considering the realisations of these three processes as time series, they might be described by their statistical features, such as half of the ratio of the mean square successive difference to the variance, $\mathcal{A}$, and the number of turning points, $T$. This paper investigates the relationships between $\mathcal{A}$ and $H$, and between $T$ and $H$. It is found numerically that the formulae $\mathcal{A}(H)=a{\rm e}^{bH}$ in case of fBm, and $\mathcal{A}(H)=a+bH^c$ for fGn and DfGn, describe well the $\mathcal{A}(H)$ relationship. When $T(H)$ is considered, no simple formula is found, and it is empirically found that among polynomials, the fourth and second order description applies best. The most relevant finding is that when plotted in the space of $(\mathcal{A},T)$, the three process types form separate branches. Hence, it is examined whether $\mathcal{A}$ and $T$ may serve as Hurst exponent indicators. Some real world data (stock market indices, sunspot numbers, chaotic time series) are analyzed for this purpose, and it is found that the $H$'s estimated using the $H(\mathcal{A})$ relations (expressed as inverted $\mathcal{A}(H)$ functions) are consistent with the $H$'s extracted with the well known wavelet approach. This allows to efficiently estimate the Hurst exponent based on fast and easy to compute $\mathcal{A}$ and $T$, given that the process type: fBm, fGn or DfGn, is correctly classified beforehand. Finally, it is suggested that the $\mathcal{A}(H)$ relation for fGn and DfGn might be an exact (shifted) $3/2$ power-law.
\end{abstract}


\end{frontmatter}

%\linenumbers

\section{Introduction}\label{intro}

Long-range dependence (LRD) is a feature of a time series that has found applications in a variety of fields, such as condensed matter \cite{lam,ferreira}, physiology \cite{gieraltowski}, Solar physics \cite{shaikh,suyal}, financial analyses \cite{vandewalle,carbone}, and astrophysical phenomena \cite{maclachlan,tarnopolski}, among others. In general, a process has the LRD property if the autocorrelations $\rho (k)$ decay to zero so slowly that their sum does not converge \cite{beran}. More specifically, when
\begin{equation}
\rho (k)\approx c_{\rho}|k|^{-\delta},
\label{eqA}
\end{equation}
where $c_{\rho}$ is a positive constant and $0<\delta<1$, the process has LRD. The dependence between events (observations) that are far apart diminishes very slowly (slower than $|k|^{-1}$) with increasing $k$. A stationary process with autocorrelations decaying as in Eq.~(\ref{eqA}) is called a stationary process with LRD or with long-term memory.

The LRD is usually quantified with the Hurst exponent \cite{hurst}, denoted $H$. To connect LRD with $H$, one introduces self-similar processes. A (stochastic) process $x(t)$ is called self-similar with a Hurst exponent $H$ if
\begin{equation}
x(t)\stackrel{\textbf{\textrm{.}}}{=}\lambda^{-H}x(\lambda t),
\label{eqB}
\end{equation}
where $\stackrel{\textbf{\textrm{.}}}{=}$ denotes equality in distribution. If the increments $x(t)-x(t-1)$ are stationary, e.g. a fractional Brownian motion (see further in the text), the autocorrelation function is given by \cite{beran}
\begin{equation}
\rho(k)=\frac{1}{2}\left( |k+1|^{2H}-2|k|^{2H}+|k-1|^{2H} \right).
\label{eqC}
\end{equation}
A Taylor expansion of $\rho(k)$ from Eq.~(\ref{eqC}) gives
\begin{equation}
\frac{\rho (k)}{H(2H-1)|k|^{2H-2}}\rightarrow 1
\label{eqD}
\end{equation}
for $k\rightarrow \infty$. It follows that for $H>1/2$, the autocorrelation $\rho(k)$ behaves like $|k|^{2-2H}$, thus $x(t)$ has LRD. 

The Hurst exponent, introduced by H. E. Hurst in 1951 \cite{hurst} to model statistically the cycle of Nile floods, is closely related to the concept of a Brownian motion (Bm, also called a random walk or a Wiener process) \cite{mandel68}, for which the consecutive increments are independent, and the standard deviation $\sigma$ scales with step $n$ as $\sigma\propto n^{1/2}$. Hurst found that in the case of Nile the increments were not independent, but characterized by a power law with an exponent greater than $1/2$. This leads to the concept of a fractional Brownian motion (fBm) \cite{mandel68}, in which the system possesses the LRD or long-term memory (also called persistency), meaning that its past increments influence the future ones, and the process tends to maintain the increments' sign. For instance, in a persistent process, if some measured quantity attains relatively high values, the system prefers to keep them high. The process is, however, probabilistic \cite{grech2}, and hence at some point the observed quantity will eventually drop to oscillate around some relatively low value. But the process still has long-term memory (being a global feature), therefore it prefers to stay at those low values until the transition occurs stochastically again. The scaling of the standard deviation in such a process is $\sigma\propto n^H$. Finally, an fBm is a non-stationary process.

$H$ can be also smaller than $1/2$. In this case, the process is anti-persistent, and it possesses short-term memory, meaning that the observed values of some quantity frequently switch from relatively high to relatively low values (more precisely speaking, the autocorrelations $\rho(k)$ decay fast enough so that their sums converge to a finite value), and there is no preference among the increments. Because the Hurst exponent is also related to the autocorrelation of a time series, i.e. to the rate of its decrease, a persistent process with $H>1/2$ is sometimes also called correlated, and an anti-persistent one, with $H<1/2$, is called anti-correlated. Finally, $H$ is bounded to the interval $(0,1)$. The properties of $H$ can be summarized as follows:
\begin{enumerate}
\item $0<H<1$,
\item $H=1/2$ for a Bm (random walk),
\item $H>1/2$ for a persistent (long-term memory, correlated) process,
\item $H<1/2$ for an anti-persistent (short-term memory, anti-correlated) process.
\end{enumerate}
Furthermore, the Hurst exponent is related to a fractal dimension of a one-dimensional time series $D\in(1,2)$ via $D=2-H$ \cite{mandel83}. This can be also generalized to processes in higher dimensions $d$: $D=d+1-H$ \cite{carbone2}. 

The scaling described above is not unique for fBm (a non-stationary process) only. It occurs also in a fractional Gaussian noise (fGn, for which $\rho (k)$ is also in the form given by Eq.~(\ref{eqC})), defined by
\begin{equation}
fGn_H(t)=B_H(t+1)-B_H(t),
\label{eq1}
\end{equation}
where $B_H$ is an fBm with a Hurst exponent equal to $H$. The increments of an fBm, forming an fGn, are described by the same $H$ as the fBm itself \cite{clegg}. Similarly, one defines a differentiated fGn (DfGn) as consecutive increments of an fGn, constructed similarly to Eq.~(\ref{eq1}), i.e. as
\begin{equation}
DfGn_H(t)=fGn_H(t+1)-fGn_H(t).
\label{eq2}
\end{equation}
Both fGn and DfGn are stationary processes.

\begin{figure}
\includegraphics[width=\columnwidth]{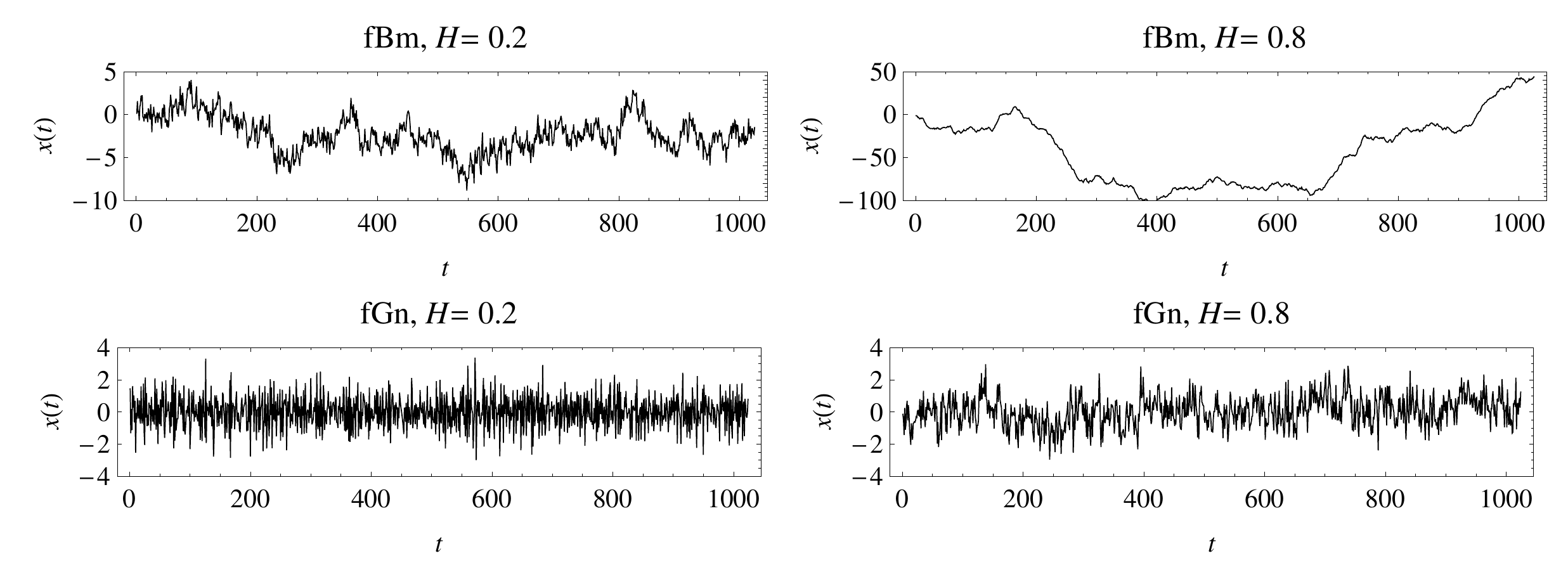}
\caption{Simulated realisations of fBm and fGn with $H=0.2$ and $H=0.8$.}
\label{fig0}
\end{figure}

Fig.~\ref{fig0} shows simulated paths of length $2^{10}$, being realisations of fBm and fGn, with $H=0.2$ and $H=0.8$, using the method described in Section~\ref{meth}. For the smaller $H$, the paths are more ragged and the range on the vertical axis is much smaller in the case of fBm due to the  reverting behavior of the time series \cite{beran}.

One can relate $H$ to the index of a power-law spectrum, $P(f)\propto 1/f^{\beta}$, where $\beta$ is a real constant for each noise type\footnote{E.g., for white noise $\beta=0$, for pink noise $\beta=1$, for blue noise $\beta=-1$. Noise with $\beta=-2$ is called a violet noise, while red (called also Brownian) noise is characterized by $\beta=2$. Noise with $\beta>2$ is called black noise. There are of course other types of noise that do not have a power-law spectrum, e.g. grey noise.}. However, it is crucial to note that this relation is different for various types of noise:
\begin{itemize}
\item $H=\frac{\beta-1}{2}$ for an fBm,
\item $H=\frac{\beta+1}{2}$ for an fGn,
\item $H=\frac{\beta+3}{2}$ for an DfGn.
\end{itemize}
This means that one has to first classify a time series under examination in order to compute the value of $H$ correctly (see Section~\ref{meth1}). The range of $\beta$ is governed by the range of the related $H\in(0,1)$.

The aim of this work is to numerically investigate the relationship between $H$ and half of the ratio of the mean square successive difference to the variance \cite{neumann}, denoted $\mathcal{A}$, and between $H$ and the number of turning points $T$ \cite{kendall,brockwell} in a time series, with an eventuality of using the latter two features to retrieve the value of $H$. This paper is organized in the following manner. In Section~\ref{meth} the methods, i.e. generating the fBm, fGn and DfGn, and retrieving the Hurst exponent, $\mathcal{A}$ and $T$, are described, and the benchmark testing is performed. In Section~\ref{res} the main results are presented. Section~\ref{ex} is devoted to conducting numerical experiments on some real-world data in order to check the applicability of the results obtained, and is followed by discussion and concluding remarks gathered in Section~\ref{conc}.

\section{Methodology}\label{meth}

Throughout this paper the computer algebra system {\sc Mathematica}\textsuperscript{\textregistered} v10.0.3 is used. The fBm and fGn for a given $H$ are generated using the built-in functionalities (the commands \texttt{FractionalBrownionMotionProcess} and\\ \texttt{FractionalGaussianNoiseProcess} are employed, respectively). The DfGn is constructed from an fGn according to Eq.~(\ref{eq2}).

\subsection{Hurst exponent}\label{meth1}

Among many existing computational algorithms for estimating $H$ (Rescaled Range Analysis ($R/S$) \cite{suyal,hurst,mandel68,mandel69}, Detrended Fluctuation Analysis (DFA) \cite{peng94,peng95,grech,hu}, wavelet approach \cite{maclachlan,tarnopolski,masry,simonsen}, Detrended Moving Average (DMA) \cite{carbone3,carbone4,carbone5}, to mention only a few), herein their extraction was performed by the wavelet approach with the Haar wavelet as a basis. The Haar wavelet was chosen due to its simplicity, most compact support and equivalence to the Allan variance \cite{xizheng}.

$H$ describes the level of statistical self-similarity of a time series. Let us recall that a time series $x(t)$ with $n$ data points is called self-similar (or self-affine) with a Hurst exponent equal to $H$ if, after a rescaling $t\rightarrow\lambda t$, it satisfies the relation in Eq.~(\ref{eqB}). The basis is obtained from a mother wavelet, $\psi (t)$, by \cite{masry}
\begin{equation}
\psi(t)\rightarrow\psi_{j,k}(t)=2^{-j/2}\psi\left(2^{-j}t-k\right),
\label{eq4}
\end{equation}
where $j$ represents the octave (time-scale) and $k$ the position of the wavelet. The wavelet transform coefficient is defined as
\begin{equation}
d_{j,k}=\braket{x,\psi_{j,k}}.
\label{eq5}
\end{equation}
The average power of a time series at a time-scale $j$ is expressed in terms of the variance of the wavelet coefficients as \cite{abry}
\begin{equation}
{\rm var}(d_{j,k})=\frac{1}{2^j}\sum\limits_{k=0}^{2^j-1}|d_{j,k}|^2.
\label{eq6}
\end{equation}
Assuming that the time series is of length $n=2^m$, $m\in\mathbb{N}$, then $2^j$ in the above formula is the number of coefficients at scale $j$ (with an intention of using the Haar wavelets; in general, $2^j$ should be replaced with an appropriate number of coefficients $n_j$). Moreover, $0\leq j\leq m-1$ and $0\leq k\leq 2^j-1$ \cite{mallat,percival}.

By inserting Eq.~(\ref{eqB}) into Eq.~(\ref{eq5}) one gets the relation between the variance of the wavelet coefficients $d_{j,k}$ and the scale $j$ \cite{flandrin}:
\begin{equation}
\log_2 {\rm var}(d_{j,k})=\alpha\cdot j+{\rm const.}
\label{eq7}
\end{equation}
The Hurst exponent is obtained by fitting a line to the linear part of the $\log_2 {\rm var}(d_{j,k})$ vs. $j$ relation in order to obtain the slope $\alpha$. The relation between the slope of the $\log_2 {\rm var}(d_{j,k})$ dependence on $j$ and $H$ is similar to the one for the power spectrum $P(f)$; here, the variances contain the information about the power carried at each time-scale $j$. Depending on the slope $\alpha$, $H$ is extracted with the following formulae:
\begin{itemize}
\item $H=\frac{\alpha-1}{2}$ when $\alpha\in(1,3)$ -- for an fBm,
\item $H=\frac{\alpha+1}{2}$ when $\alpha\in(-1,1)$ -- for an fGn,
\item $H=\frac{\alpha+3}{2}$ when $\alpha\in(-3,-1)$ -- for a DfGn.
\end{itemize}

Contrary to the other mentioned methods, the wavelet approach allows to classify the examined time series depending on the obtained slope $\alpha$, it is relatively faster in numerical implementations, and was found to be more robust for time series for which the LRD is not as equivocal (see Sect.~\ref{ex}). For further details, the reader is referred, e.g., to \cite{maclachlan,tarnopolski,peng94,peng95}.

\subsection{Ratio of the mean square successive difference to the variance}\label{meth2}

The statistical properties of twice the Abbe value \cite{kendall1971}, denoted $\mathcal{A}$, were examined in detail in \cite{neumann} (see also \cite{neumann2}). For a time series $\{x_i\}_{i=1}^n$, $\mathcal{A}$ is defined as
\begin{equation}
\mathcal{A}=\frac{\delta^2}{2s^2}=\frac{\frac{1}{n-1}\sum\limits_{i=1}^{n-1}(x_{i+1}-x_i)^2}{\frac{2}{n}\sum\limits_{i=1}^n (x_i-\bar{x})^2},
\label{eq8}
\end{equation}
where $\bar{x}$ is the mean of $\{x_i\}$. The Abbe value quantifies the smoothness of a time series by comparing the sum of the squared differences between two successive measurements with the standard deviation of the time series. It decreases to zero for time series displaying a high degree of smoothness, while the normalization factor in Eq.~(\ref{eq8}) ensures that $\mathcal{A}$ tends to unity for a purely noisy time series (more precisely, white noise) \cite{williams}. The quantity $\mathcal{A}$, as defined, is a random variable.

\subsection{Turning points analysis}\label{meth3}

Let us consider three consecutive observations in a time series, $x_{i-1},x_i,x_{i+1}$; assume that $x_j\neq x_k$ for any two of them. These measurements can be arranged in six ways; in four of them the middle value will be greater or lower than the two that surround it. This configuration is called a turning point \cite{kendall,brockwell}. A probability of finding a turning point at time $i$ is equal to $2/3$ for a random time series. Denote the number of turning points in a time series by $T$; as defined it is a random variable. Having a random time series of length $n$, the expected value of $T$ is given by
\begin{equation}
\mu_T=\frac{2}{3}(n-2),
\label{eq10}
\end{equation}
because the first and last measurements ($x_1$ and $x_n$) cannot form a turning point. For a purely noisy (random) time series, $\mu_T$ tends to $2n/3$. In what follows the ratio $T/\mu_T$ will be considered, hence a white noise is characterized by $T/\mu_T=1$. A time series with $T/\mu_T>1$ (i.e., with raggedness exceeding the one of a Gaussian noise) will be more noisy than white noise. Similarly, a time series with $T/\mu_T<1$ will be ragged less than white noise.

\subsection{Hurst exponent benchmark testing}\label{meth4}

\begin{figure}
\centering
\includegraphics[width=0.5\columnwidth]{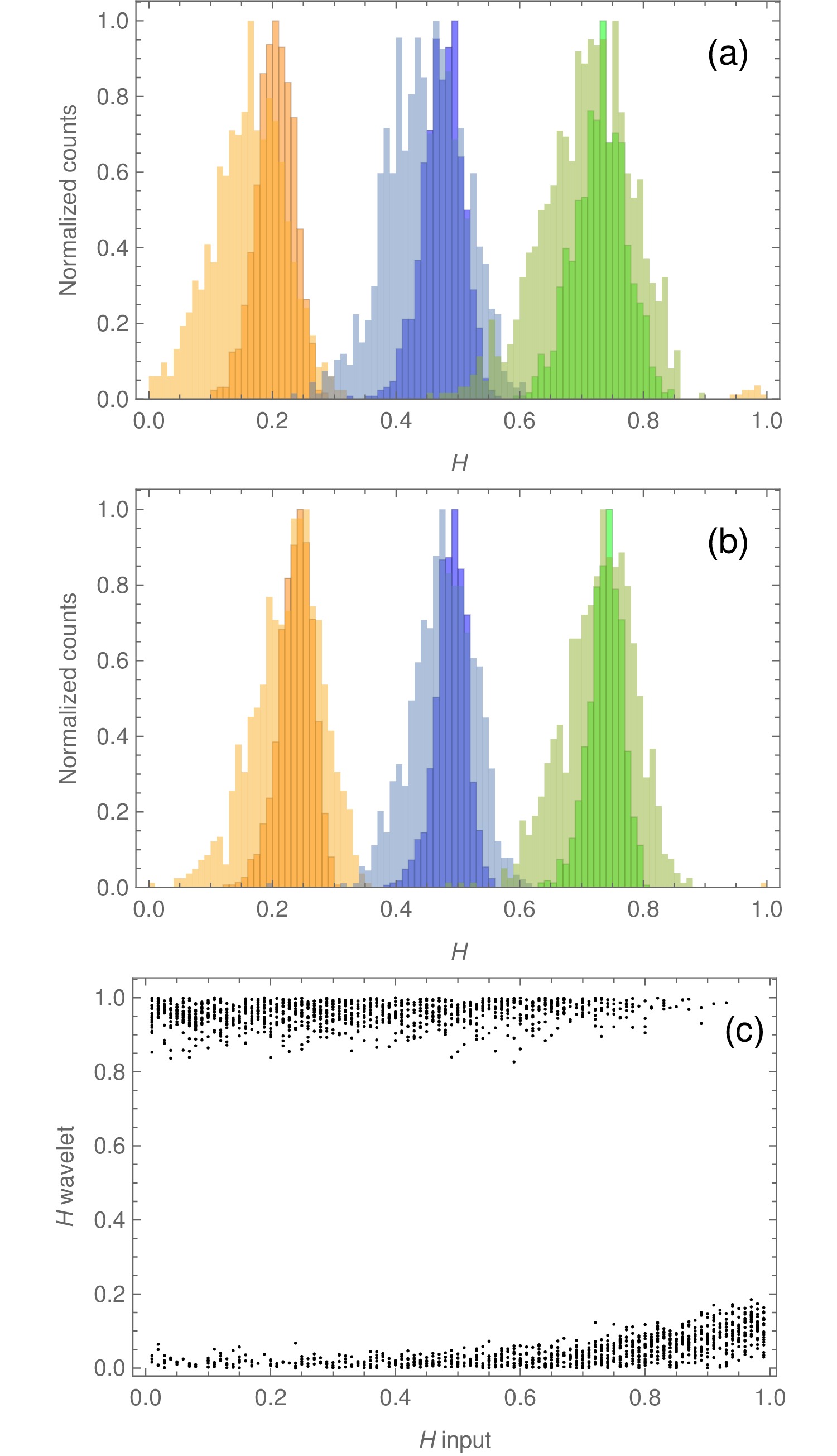}
\caption{(a) and (b) Histograms of simulated fBm and fGn realisations, respectively. For each process type, $N=1000$ simulations were performed for $H=0.25$, $0.50$ and $0.75$. These time series were used to recover $H$ using the wavelet method as described in Section~\ref{meth1}. The lighter histograms with $n=2^{10}$ are overlaid with darker ones that were obtained for $n=2^{14}$. In (c) the results for DfGn are displayed for $H\in[0.01,0.99]$, in steps of $\Delta H=0.01$, and $n=2^{12}$.}
\label{fig1}
\end{figure}

The algorithm for computing $H$, described in Section~\ref{meth1}, does not only return the value of $H$, but also the type of the underlying process, based on the slope $\alpha$. In order to estimate its reliability for the three types of processes examined herein (fBm, fGn, and DfGn), the following procedure is undertaken. For each process type, 1000 realisations of a time series are produced for each $H=0.25,\,0.50,\,0.75$. This is performed for two lengths: $n=2^{10}$ and $n=2^{14}$. Figures~\ref{fig1} (a) and (b) display the results in form of histograms of the attained values of $H$ for fBm and fGn. It is visible that the algorithm slightly underestimates the Hurst exponent, and the smaller the $H$ is, the bigger the underestimation. Nevertheless, the standard deviations of the data (see Table~\ref{tbl1}) convince that the method is trustworthy enough for usual applications (compare also with \cite{maclachlan}).

\begin{table}
\caption{Summary of results in Fig.~\ref{fig1} (a) and (b); the mean is the sample mean, and std stands for standard deviation of the sample, with the sample size $N=1000$.}
\label{tbl1}
\centering
\begin{tabular}{ccccc}
\hline\hline
Process & $n$ & $H_{\rm input}$ & mean & std \\
\hline
    &			& 0.25 & 0.17 & 0.11 \\
fBm &  $2^{10}$ & 0.50 & 0.45 & 0.06 \\
    &			& 0.75 & 0.71 & 0.07 \\
\hline
    &			& 0.25 & 0.20 & 0.03 \\
fBm &  $2^{14}$ & 0.50 & 0.47 & 0.03 \\
    &			& 0.75 & 0.73 & 0.04 \\
\hline
    &			& 0.25 & 0.23 & 0.07 \\
fGn &  $2^{10}$ & 0.50 & 0.47 & 0.06 \\
    &			& 0.75 & 0.73 & 0.05 \\
\hline
    &			& 0.25 & 0.24 & 0.03 \\
fGn &  $2^{14}$ & 0.50 & 0.49 & 0.03 \\
    &			& 0.75 & 0.74 & 0.03 \\
\hline
\end{tabular}
\end{table}

However, as displayed in the scatter plot in Figure~\ref{fig1} (c), it is strongly biased in the case of DfGn: the resultant $H$'s gather at either high ($\gtrsim 0.8$), or low values ($H\lesssim 0.2$). Also, the classification of the time series was incorrect for 943 out of 1980 cases. Hence, while the wavelet method is efficient for fBm and fGn, it is not suitable in general for DfGn-type time series. A solution to this issue is to convert a DfGn into an fGn by means of cumulative sums, i.e. to invert Eq.~(\ref{eq2}).

\section{Results}\label{res}

\begin{figure}
\includegraphics[width=\columnwidth]{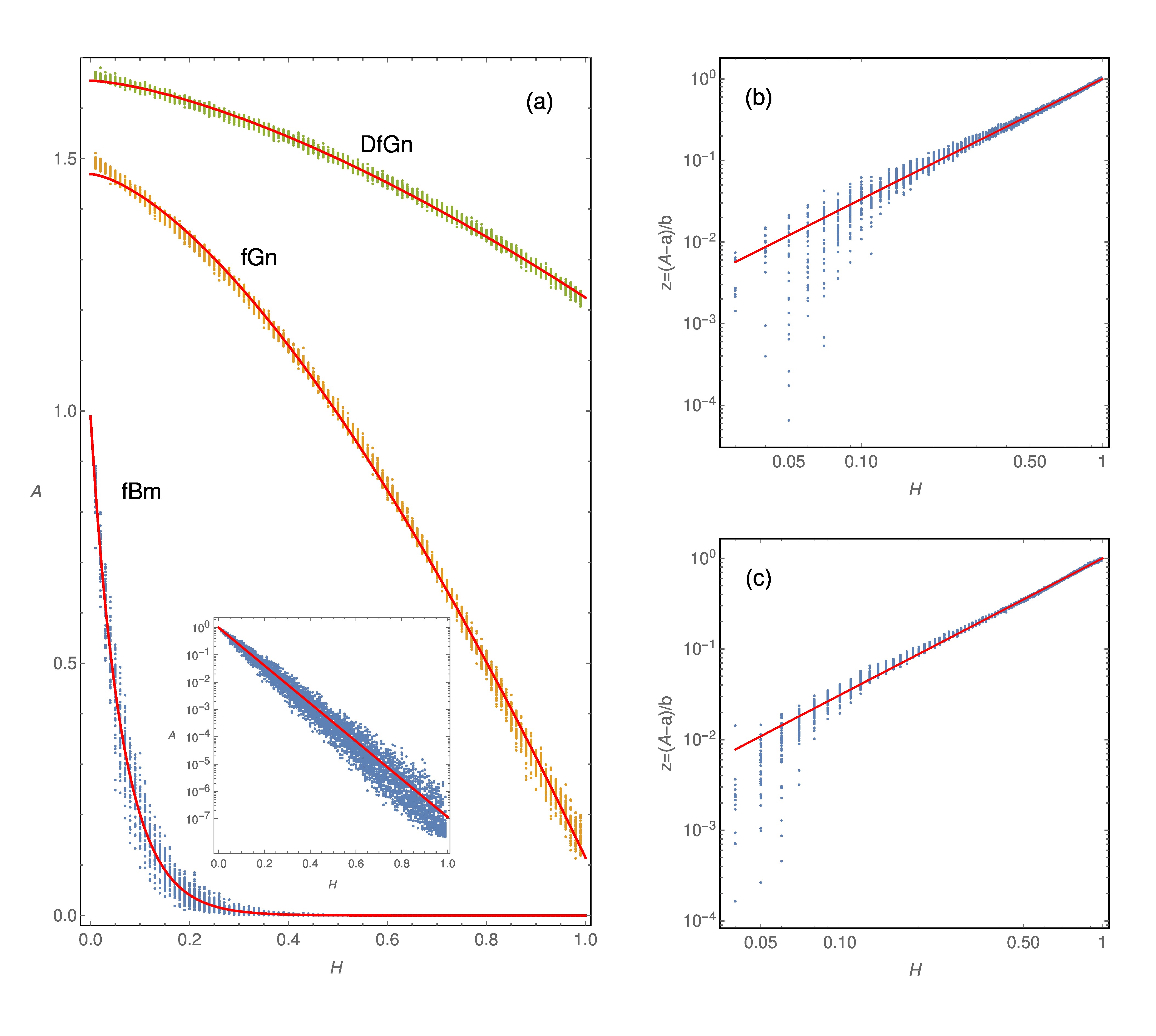}
\caption{(a) The relation between the Abbe value and the Hurst exponent, for $H\in[0.01,0.99]$, in steps of $\Delta H=0.01$, and with 50 realisations for each $H$ value, for each process type. The red lines are the fitted functions: $\mathcal{A}=b\cdot \exp{(aH)}$ for fBm, and $\mathcal{A}=a+bH^c$ for fGn and DfGn. The fits for fGn and DfGn deviate slightly near $H\approx 0$. The inset shows the $\mathcal{A}$ vs. $H$ relation in a semi-log plot. (b) and (c) show the rescaled Abbe value, $z=(\mathcal{A}-a)/b$, versus $H$ in a log-log diagram, for the fGn and DfGn, respectively.}
\label{fig2}
\end{figure}

\begin{figure}
\includegraphics[width=\columnwidth]{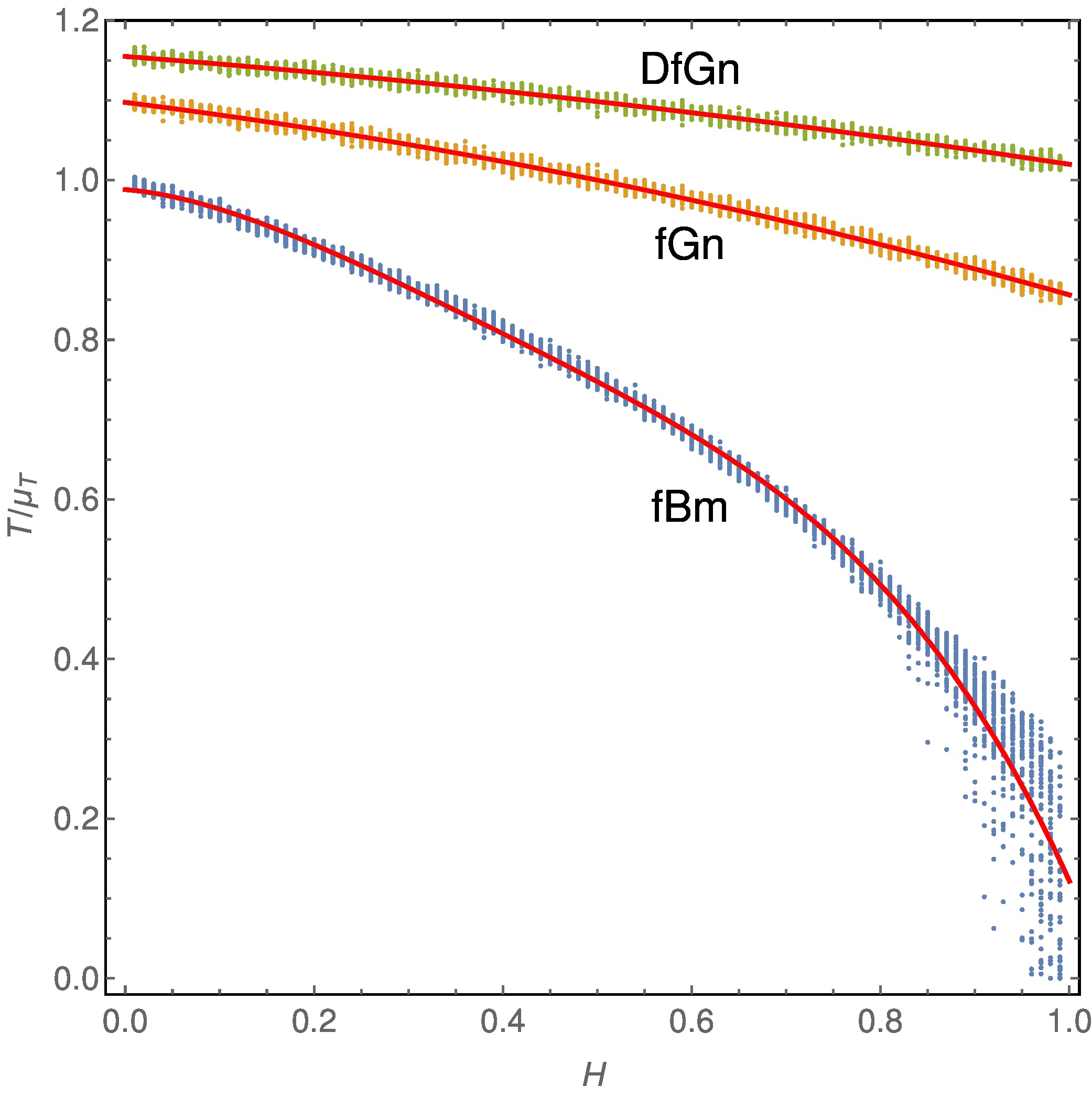}
\caption{The relation between the ratio $T/\mu_T$ and $H$, for $H\in[0.01,0.99]$, in steps of $\Delta H=0.01$, and with 50 realisations for each $H$ value, for each process type. The red lines are the fitted polynomials, parameters of which are gathered in Table~\ref{tbl3}.}
\label{fig3}
\end{figure}

To investigate the relation between $H$, $\mathcal{A}$ and $T/\mu_T$, 4950 realisations were generated for fBm, fGn and DfGn each, for $H\in[0.01,0.99]$, in steps of $\Delta H=0.01$, and $n=2^{14}$. Next, the Abbe value given by Eq.~(\ref{eq8}), and the ratio $T/\mu_T$ as described in Section~\ref{meth3}, were computed. The results, in form of a scatter plot, are displayed in Figs.~\ref{fig2} and \ref{fig3}. The inset in Fig.~\ref{fig2} (a), showing the relation between $\mathcal{A}$ and $H$ for fBm on a semi-log plot, indicates that the formula $\mathcal{A}(H)=a{\rm e}^{bH}$ holds. Fig.~\ref{fig2} (b) and (c) convince that for fGn and DfGn the functional relation in form of a shifted power-law, $\mathcal{A}(H)=a+bH^c$, is a good description. Here, $a,b,c$ are real constants. Parameters of the fits, obtained by means of the least-squares fitting, are gathered in Table~\ref{tbl2}. The standard errors confirm that the exponential (for fBm) or shifted power-law (for fGn and DfGn) relation $\mathcal{A}(H)$ is indeed a good guess. The choice of these particular functions is a result of trial and error using relatively simple elementary functions with as little free parameters as possible (a {\sc Mathematica}\textsuperscript{\textregistered} v10.0.3 built-in command \texttt{FindFormula} was used for guidance).
\begin{table}
\caption{Parameters of the fits of the relation $\mathcal{A}(H)$ for $n=2^{14}$. The errors in brackets correspond to the last significant digit.}
\label{tbl2}
\centering
\begin{tabular}{ccccc}
\hline\hline
Process & Formula & $a$ & $b$ & $c$ \\
\hline
fBm & $\mathcal{A}(H)=a{\rm e}^{bH}$ & $0.989(2)$ & $-15.95(5)$ & --- \\
fGn & $\mathcal{A}(H)=a+bH^c$ & $1.4695(5)$ & $-1.3550(7)$ & $1.508(2)$ \\
DfGn & $\mathcal{A}(H)=a+bH^c$ & $1.6546(3)$ & $-0.4300(3)$ & $1.474(3)$ \\
\hline
\end{tabular}
\end{table}

The functional relation between $\mathcal{A}$ and $H$ is different for fBm, and for fGn and DfGn, what might be surprising at first in light of the simple differential relationships between the three process types examined. It is worth to note, however, that fBm is a non-stationary process, while both fGn and DfGn are stationary processes, hence their different statistical properties may influence the functional relation between $\mathcal{A}$ and $H$. Due to a numerical approach undertaken herein, it is not straightforward to verify whether the stationarity and non-stationarity of the processes is an explanation sufficient for the different functional forms. Nevertheless, it might be a plausible reason, or at least a hint towards a more justified explanation to be given in the future.

Analysing Fig.~\ref{fig2} (a) the following observations are formulated:
\begin{itemize}
\item if $\mathcal{A}\lesssim 0.1$, then the process is fBm-like;
\item if $0.1\lesssim\mathcal{A}\lesssim1$, then the process is fBm-like or fGn-like;
\item if $1\lesssim\mathcal{A}\lesssim 1.2$, then the process is fGn-like;
\item if $1.2\lesssim\mathcal{A}\lesssim 1.5$, then the process is fGn-like or DfGn-like;
\item if $\mathcal{A}\gtrsim 1.5$, then the process is DfGn-like.
\end{itemize}
Because the ranges of $\mathcal{A}$ overlap rather significantly (however, $\mathcal{A}\lesssim 0.1$ corresponds to $H\gtrsim 0.14$ in case of fBm), the Abbe value by itself cannot be unambiguously used to infer the value of the corresponding $H$. Hence, the dependence of the ratio $T/\mu_T$ on $H$ is also investigated, and the results are displayed in Fig.~\ref{fig3}. It follows that
\begin{itemize}
\item if $T/\mu_T\lesssim 0.85$, then the process is fBm-like;
\item if $0.85\lesssim T/\mu_T\lesssim1$, then the process is fBm-like or fGn-like;
\item if $1\lesssim T/\mu_T\lesssim 1.1$, then the process is fGn-like or DfGn-like;
\item if $T/\mu_T\gtrsim 1.1$, then the process is DfGn-like.
\end{itemize}
Hence, the ranges of $T/\mu_T$ also overlap significantly for the whole range of $H$.

Because the functional form of the relation is not as obvious as it was for $\mathcal{A}(H)$, polynomials of various degree are fitted to the obtained dependencies. The parameters are gathered in Table~\ref{tbl3}. The plynomial's degree for each process is chosen as the one at which the $\chi^2$ of the fits plateau. Obviously the polynomial forms are certainly data-driven and are presented here only for the sake of completeness.
\begin{table}
\caption{Parameters of the fits of the polynomial dependence of $T/\mu_T$ on $H$.}
\label{tbl3}
\centering
\begin{tabular}{cc}
\hline\hline
Process & Formula \\
\hline
fBm  & $-1.94(8)H^4+2.91(16)H^3-1.73(11)H^2-0.09(3)H+0.988(1)$ \\
fGn  & $-0.093(1)H^2-0.148(1)H+1.0974(2)$ \\
DfGn & $-0.0441(9)H^2-0.0910(9)H+1.1549(2)$ \\
\hline
\end{tabular}
\end{table}

\begin{figure}
\includegraphics[width=\columnwidth]{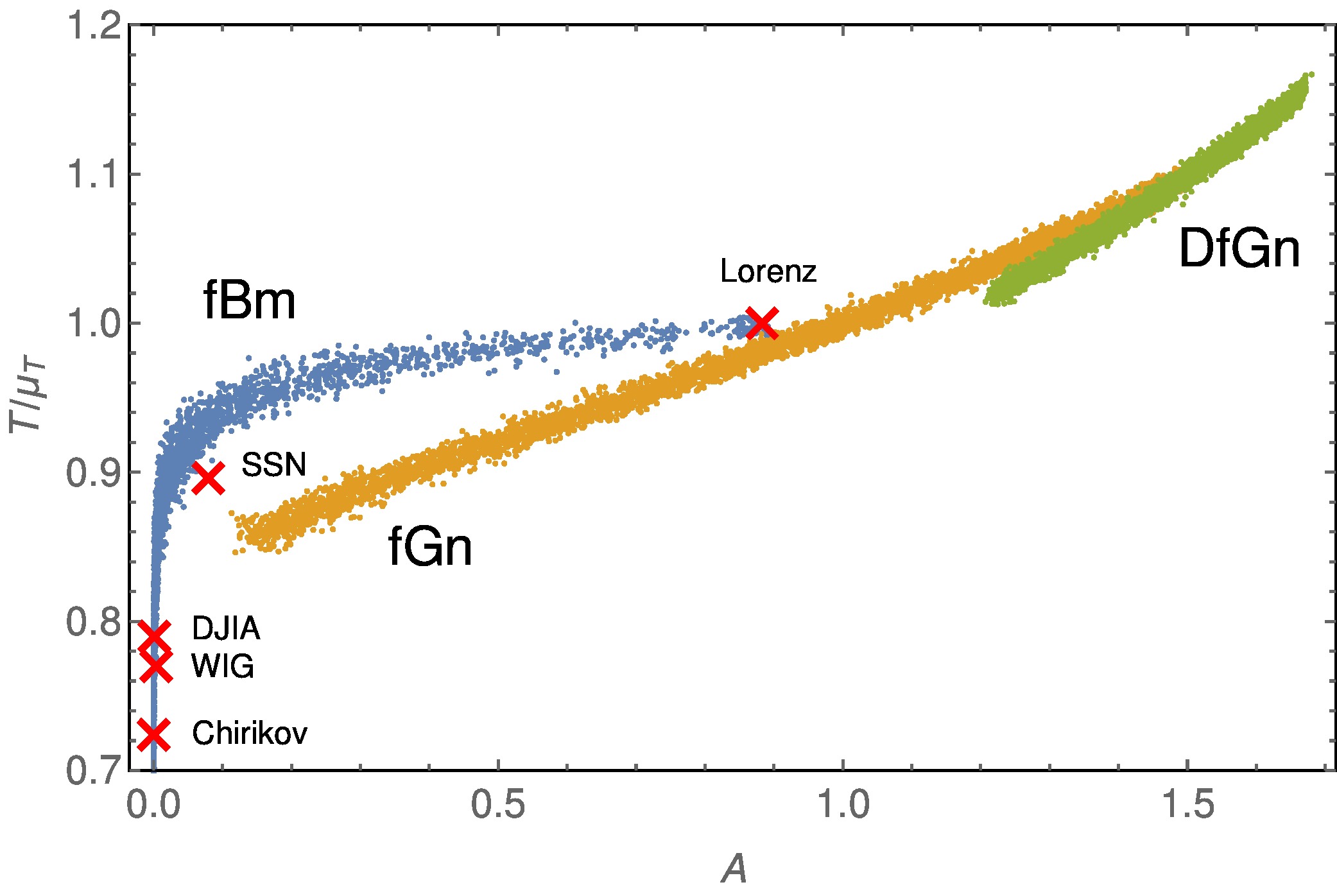}
\caption{The relation between the ratio $T/\mu_T$ and the Abbe value for fBm, fGn and DfGn, and $n=2^{14}$. The fBm branch extends below $T/\mu_T$ nearly vertically until it reaches zero. The red crosses mark the locations of the time series examined in Sect.~\ref{ex}. See text for details.}
\label{fig4}
\end{figure}

Despite the dependencies of $\mathcal{A}$ and $T/\mu_T$ on $H$ being generally not unambiguous, as their values overlap in some of their ranges for all $H$, the three processes might be separated in a space of $T/\mu_T$ vs. $\mathcal{A}$. To verify this proposition, the obtained pairs $(\mathcal{A},T/\mu_T)$ are shown in a scatter plot in Fig.~\ref{fig4}, and indeed fBm, fGn and DfGn are clearly distinguishable. The distinction is particularly sharp between the fBm and fGn; the DfGn slightly overlaps with the right-hand side of the fGn branch.

\section{Numerical examples}\label{ex}

In order to check the validity and usefulness of the relationships discovered in Sect.~\ref{res}, the following time series are examined:
\begin{itemize}
\item a time series of the $x$ variable of the Lorenz system \cite{lorenz}
\begin{equation}
\dot{x}=\sigma (y-x),\quad\dot{y}=x(\tau-z)-y,\quad\dot{z}=xy-\beta z,
\label{eq11}
\end{equation}
with $n=2^{14}$ and a time step of $\Delta t=1$. The initial conditions are $(x_0,y_0,z_0)=(1,5,10)$, and the standard parameters are applied: $(\sigma,\tau,\beta)=(10,28,8/3)$. This system was chosen due to a claim \cite{suyal} that it exhibits persistent behavior;
\item the last $2^{11}$ values of the monthly mean of the total sunspot number (SSN), obtained from \url{http://www.sidc.be/silso/datafiles} (source: WDC-SILSO, Royal Observatory of Belgium, Brussels; accessed on October 31, 2015). The Hurst exponent is widely used to examine whether the Solar activity variations might be caused by a random (i.e., white-noise) process \cite{shaikh,suyal};
\item the last $2^{11}$ values of the Dow Jones Industrial Index (DJIA), obtained from \url{https://research.stlouisfed.org/fred2/series/DJIA/downloaddata} (accessed on October 31, 2015);
\item the last $2^{10}$ values of the Warsaw Stock Exchange Index (Warszawski Indeks Gie\l dowy, WIG), obtained from \url{http://www.investing.com/indices/wig-historical-data} (accessed on October 31, 2015). The Hurst exponent has long been applied to financial time series in order to infer any long-term trends in their behavior and to possibly forecast their future values \cite{grech2,tzouras};
\item a time series of the $p$ variable of the Chirikov standard map \cite{chirikov,lichtenberg}
\begin{equation}
p_{n+1}=p_n+\frac{K}{2\pi}\sin(2\pi x_n),\quad x_{n+1}=x_n+p_{n+1},
\label{eq12}
\end{equation}
with $n=2^{14}$ taken after discarding the initial $2^{14}$ values to ensure that transient behavior is skipped over. The initial conditions were chosen so that a reference to other results \cite{manchein,tarnopolski2} is straightforward: $(p_0,x_0)=(0,0.2865)$, and $K=3.228259$.
\end{itemize}

\begin{figure}
\includegraphics[width=\columnwidth]{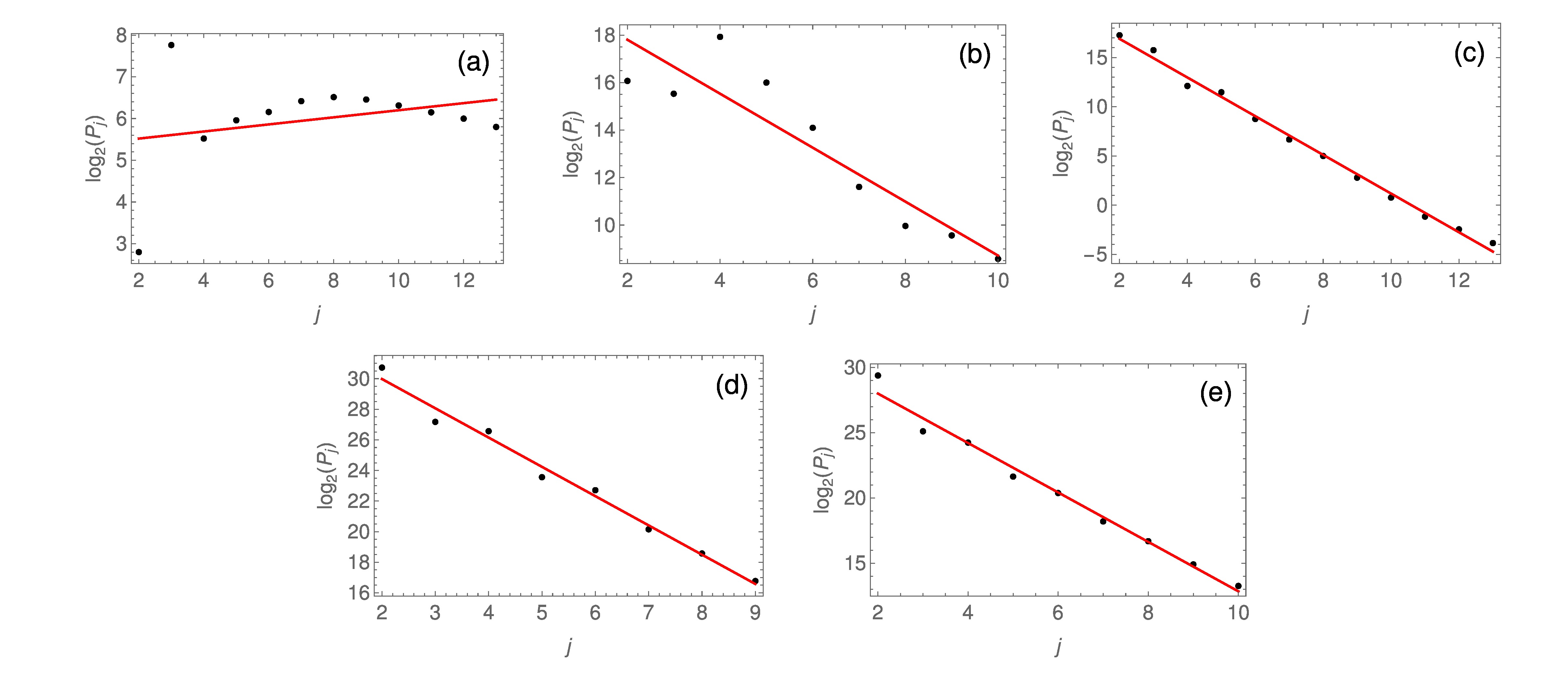}
\caption{Log-scale diagrams for the samples: (a) Lorenz system, (b) SSN, (c) Chirikov standard map, (d) WIG, and (e) DJIA.}
\label{fig5}
\end{figure}

For each of the above time series, $\mathcal{A}$ given by Eq.~(\ref{eq8}), and $T/\mu_T$ as described in Section~\ref{meth3}, were computed. The resultant values are placed in the space of $T/\mu_T$ vs. $\mathcal{A}$, and are displayed as red crosses in Fig.~\ref{fig4}. Next, $H$ is estimated using the $H(\mathcal{A})$ relations, obtained by inverting $\mathcal{A}(H)$ from Table~\ref{tbl2}. This value is compared with the Hurst exponent extracted by means of the wavelet method as described in Sect.~\ref{meth1}. A summary of this procedure is presented in Table~\ref{tbl4}. In order to illustrate the errors behind the obtained values of $H$, the linear regressions are shown in the so called log-scale diagrams in Fig.~\ref{fig5}.
\begin{table}
\caption{The ratio $T/\mu_T$, Abbe value $\mathcal{A}$, and $H$ for the tested samples. The process type is inferred based on the slope $\alpha$ of the linear regression in the $H$ estimation using the wavelet method; $H_{\rm wavelet}$ is an $H$ obtained with the wavelet approach, and $H(\mathcal{A})$ is an $H$ estimated using the appropriate (inverted) formula from Table~\ref{tbl2}. Regarding the Lorenz system, results inferred directly from the slope are marked in bold, and the other possibility is displayed as well.}
\label{tbl4}
\centering
\begin{tabular}{lcccccc}
\hline\hline
Sample & $n$ & Process type & $T/\mu_T$ & $\mathcal{A}$ & $H(\mathcal{A})$ & $H_{\rm wavelet}$ \\
\hline
Lorenz   & $2^{14}$ & \begin{tabular}{@{}c@{}}{\bf fGn}/ \\ fBm\end{tabular} & 0.999 & 0.8826 & \begin{tabular}{@{}c@{}}$\mathbf{0.5743\pm0.0006}$ \\ $0.0071\pm0.0002$\end{tabular} & --- \\
SSN      & $2^{11}$ &      fBm  & 0.895 & 0.0797 & $0.158\pm0.001$ & $0.068\pm0.187$ \\
Chirikov & $2^{14}$ &      fBm  & 0.723 & 0.0002 & $0.548\pm0.001$ & $0.483\pm0.048$ \\
WIG      & $2^{10}$ &      fBm  & 0.769 & 0.0040 & $0.346\pm0.001$ & $0.457\pm0.094$ \\
DJIA     & $2^{11}$ &      fBm  & 0.789 & 0.0012 & $0.419\pm0.001$ & $0.447\pm0.094$ \\
\hline
\end{tabular}
\end{table}

\begin{figure}
\includegraphics[width=\columnwidth]{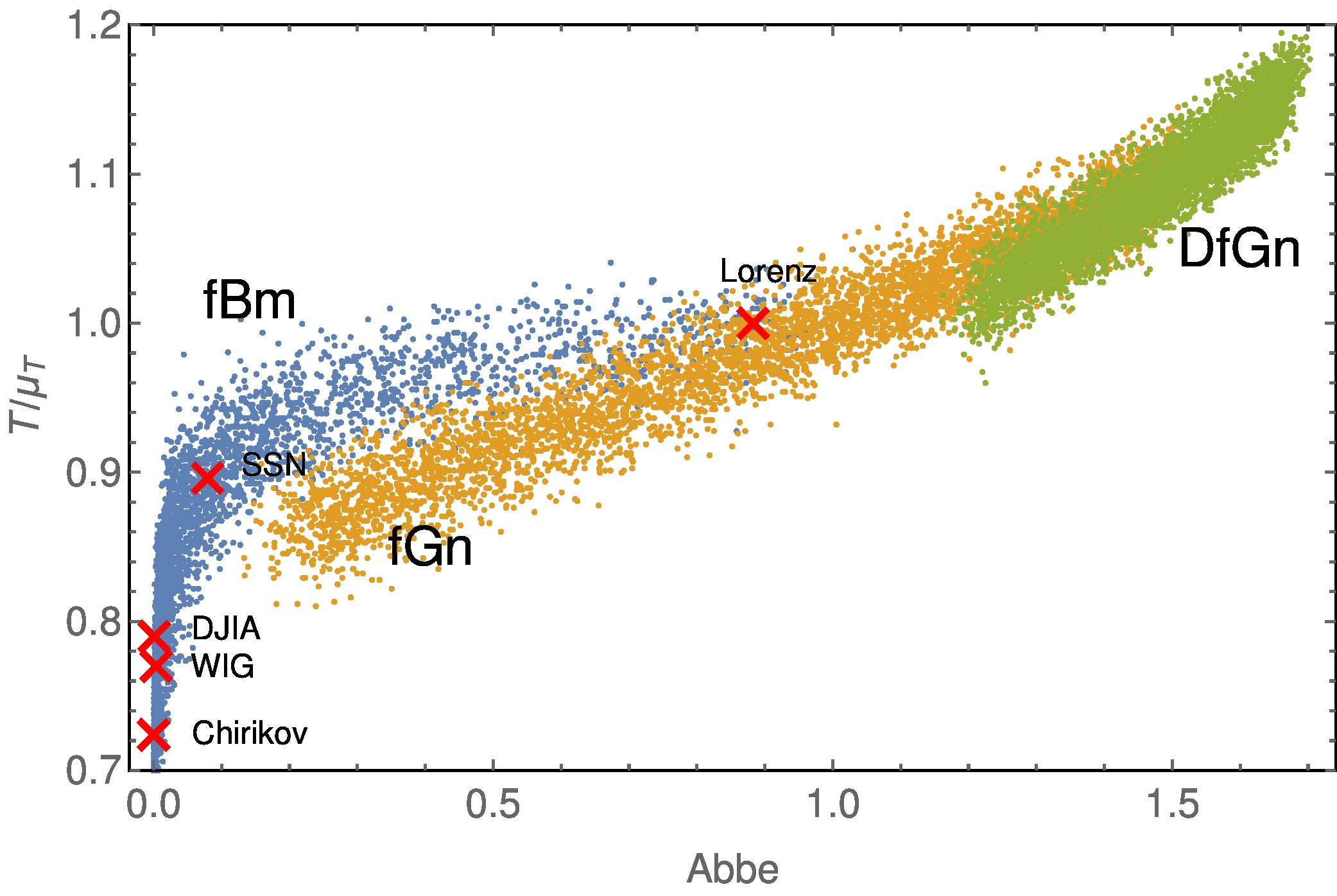}
\caption{The same as Fig.~\ref{fig4}, but for $n=2^{10}$.}
\label{fig6}
\end{figure}

While the time series of the Chirikov standard map, DJIA and WIG are located quite confidently on the fBm branch in Fig.~\ref{fig4}, the SSN time series is slightly departed from it. It has length $n=2^{11}$, as well as the DJIA. Because the WIG is the shortest dataset considered herein, Fig.~\ref{fig4} is reproduced in Fig.~\ref{fig6}, but for $n=2^{10}$. The general behavior of the $T/\mu_T$ vs. $\mathcal{A}$ relation appears to be the same, but the dispersion of the points forming the fBm, fGn and DfGn branches---i.e., the branches' widths---is significantly greater than compared to the results for $n=2^{14}$ (compare with Fig.~\ref{fig4}). Similarly, the dependencies of $T/\mu_T$ and $\mathcal{A}$ on $H$ are also wider for $n=2^{10}$ (not shown). The values of $\mathcal{A}$ and $T/\mu_T$ place the SSN well within the fBm branch in this case. However, the overlap of the fBm and fGn branches is much greater than it was for the case of $n=2^{14}$ (although SSN does not lie in the overlapping region). 

Table~\ref{tbl5} gathers the parameters of the $\mathcal{A}(H)$ relations for $n=2^{10}$. Using these values, the $H$'s were again estimated based on the Abbe values, and are displayed in Table~\ref{tbl6}. For fBms, they are systematically higher than they were for $n=2^{14}$, for which the values of $H(\mathcal{A})$ are in agreement with the $H_{\rm wavelet}$ values. For the smaller $n=2^{10}$, they are also in agreement with $H_{\rm wavelet}$, except the $H$ of the time series from the Chirikov standard map, that is greater by $0.29$ and $0.355$ than the $H(\mathcal{A})$ and $H_{\rm wavelet}$ values, respectively, obtained for $n=2^{14}$.
\begin{table}
\caption{The same as Table~\ref{tbl2}, but for $n=2^{10}$.}
\label{tbl5}
\centering
\begin{tabular}{ccccc}
\hline\hline
Process & Formula & $a$ & $b$ & $c$ \\
\hline
fBm & $\mathcal{A}(H)=a{\rm e}^{bH}$ & $0.986(3)$ & $-10.44(5)$ & --- \\
fGn & $\mathcal{A}(H)=a+bH^c$ & $1.479(1)$ & $-1.304(2)$ & $1.425(5)$ \\
DfGn & $\mathcal{A}(H)=a+bH^c$ & $1.6556(9)$ & $-0.431(1)$ & $1.46(1)$ \\
\hline
\end{tabular}
\end{table}

\begin{table}
\caption{Hurst exponents for the tested samples. The symbols are the same as in Table~\ref{tbl4}, except that $H(\mathcal{A})$ is the $H$ estimated using the appropriate inverted formula from Table~\ref{tbl5} (i.e., for $n=2^{10}$). Time series' lengths, process types and $H_{\rm wavelet}$ are displayed for self-completeness of this Table.}
\label{tbl6}
\centering
\begin{tabular}{lcccc}
\hline\hline
Sample & $n$ & Process type & $H(\mathcal{A})$ & $H_{\rm wavelet}$ \\
\hline
Lorenz   & $2^{14}$ & \begin{tabular}{@{}c@{}}{\bf fGn}/ \\ fBm\end{tabular} & \begin{tabular}{@{}c@{}}$\mathbf{0.578\pm0.002}$ \\ $0.0106\pm0.0003$\end{tabular} & --- \\
SSN      & $2^{11}$ & fBm  & $0.241\pm0.001$ & $0.068\pm0.187$ \\
Chirikov & $2^{14}$ & fBm  & $0.838\pm0.004$ & $0.483\pm0.048$ \\
WIG      & $2^{10}$ & fBm  & $0.528\pm0.002$ & $0.457\pm0.094$ \\
DJIA     & $2^{11}$ & fBm  & $0.641\pm0.003$ & $0.447\pm0.094$ \\
\hline
\end{tabular}
\end{table}

\section{Discussion and conclusions}\label{conc}

The dependencies of the Abbe value $\mathcal{A}$ and the ratio $T/\mu_T$ on $H$ were examined numerically for fBm, fGn and DfGn processes, with time series of length $n=2^{14}$, and tight relationships were found. The functional form of $\mathcal{A}(H)$ in case of an fBm, based on a semi-log plot, is quite confidently guessed to be exponential: $\mathcal{A}(H)=a{\rm e}^{bH}$. When the fGn and DfGn are considered, after a simple (linear) transformation and displaying the results in a log-log plot, it was found that a shifted power-law in the form $\mathcal{A}(H)=a+bH^c$ is a good description of the observed relations.

The ratio $T/\mu_T$ was also found to be tightly dependent on $H$, however no simple functional form appears to underly this relationship. A crude procedure of fitting polynomials of various degrees led to a fourth order polynomial dependence in case of fBm, and a second order relation was inferred for fGn and DfGn.

Finally, the three processes under consideration were found to form distinct branches in a space of $T/\mu_T$ vs. $\mathcal{A}$. This allows to classify an unknown process based on its Abbe value and the ratio $T/\mu_T$. It was further verified that the shape of the above described dependencies was roughly the same for shorter time series (with $n=2^{10}$), but with a larger scatter of the points, i.e. the branches overlap stronger than for longer time series (with $n=2^{14}$) due to their greater width. Hence, the confidence in assessing the process type based on $T/\mu_T$ and $\mathcal{A}$ depends on the number of data points gathered. It should be expected that the scatter ought to decrease for longer time series, asymptotically leading to an analytical relationships between all the pairs among $H$, $\mathcal{A}$ and $T/\mu_T$.

Because the $\mathcal{A}(H)$ relation is invertible for all three process types, it may be used to estimate the Hurst exponent as $H(\mathcal{A})$. This proposition was tested for different data (two chaotic systems: discrete and continuous, SSN, DJIA and WIG), and was in general found to give results consistent with the $H$'s obtained using the wavelet method. However, there was an ambiguity in case of the time series from the Lorenz system: being placed quite confidently on the fBm branch, it implied an $H=0.007$, but the wavelet approach did not allow to infer any value of $H$ for reference due to a lacking linear part in the log-scale diagram. On the other hand, it is located near the fGn branch, and if the formula $H(\mathcal{A})$ for fGn is used, an $H=0.57$ is obtained, which is a qualitatively different result.

Nevertheless, the $H$'s obtained by means of the $H(\mathcal{A})$ relation for the other four exemplary time series were consistent within the standard errors with the $H$'s extracted with the use of the wavelet method. Therefore, the proposed functional relationship between $H$ and $\mathcal{A}$, aided with the value of $T/\mu_T$, allows an estimation of the Hurst exponent of a time series coming from an fBm-like process that is consistent with the wavelet approach. Becasue the relationships inferred from longer realisations of fBm, fGn and DfGn (i.e., with $n=2^{14}$) are much tighter, i.e. the scatter among the points is smaller (compare Figs.~\ref{fig2}, \ref{fig3}, \ref{fig4}, \ref{fig6}), the $H$'s obtained via the phenomenological $H(\mathcal{A})$ expressions should be expected to be closer to the real values of $H$. This is justified with the $H$'s, obtained from the Abbe values, being closer to $H_{\rm wavelet}$ than the ones achieved with the use of the functional relations obtained with shorter fBm, fGn and DfGn time series.

Eventually, it should be emphasized that the rather simple and tight relation---exponential or power-law---of $\mathcal{A}$ on $H$ is a result that could not be predicted from the start. While some kind of correlation is straightforward, as the Abbe value measures the smoothness of a time series, and the Hurst exponent is a measure of its raggedness, it is quite surprising that the $\mathcal{A}(H)$ relationship is so tight and simple in its mathematical description. This allows to efficiently estimate the Hurst exponent based on fast and easy to compute $\mathcal{A}$ and $T$, given that the process type: fBm, fGn or DfGn, is correctly classified beforehand.

Unfortunately, an attempt to derive the $\mathcal{A}(H)$ formula analytically was unsuccesful. Nevertheless, it seems likely that it might be possible in the future. Finally, because the exponent in the $\mathcal{A}(H)$ relation for fGn and DfGn is nearly 1.5, it is suggested herein that it may be an exact $3/2$ power-law.

\section*{Acknowledgements}

The author acknowledges support in a form of a special scholarship of Marian Smoluchowski Scientific Consortium Matter-Energy-Future from KNOW funding, grant number KNOW/48/SS/PC/2015, and wishes to thank the anonymous reviewers for useful comments that lead to significant improvement and clarification of the paper.

\section*{References}

\bibliography{mybibfile}

\end{document}